\documentclass[12pt]{iopart}

\usepackage{iopams}  

\usepackage[utf8]{inputenc} 
\usepackage[T1]{fontenc}
\usepackage[british]{babel}

\usepackage[
  input-uncertainty-signs=\pm,
  separate-uncertainty = true,
  group-separator = {\,},
	detect-weight=true,
  detect-family,
]{siunitx}
\DeclareSIUnit[per-mode=symbol]\mps{\m\per\s\squared}

\usepackage[numbers,square,sort&compress]{natbib}
\usepackage[pdftex]{graphicx} 
\usepackage{subcaption}

\usepackage{nicefrac}
\expandafter\let\csname equation*\endcsname\relax
\expandafter\let\csname endequation*\endcsname\relax
\usepackage{amsmath}
\usepackage{bm}
\usepackage{xspace}
\usepackage{multirow}

\usepackage[version=4]{mhchem} 
\newcommand{\eg}{e.\,g.\@\xspace}
\newcommand{\vkeff}{\vec{k}_\ensuremath{\mathrm{eff}}\xspace}
\newcommand{\msqwert}[1]{\qty{#1}{\mps\per\sqrt{\text{\Hz}}}}
\newcommand{\keff}{k_\ensuremath{\mathrm{eff}}\xspace}
\usepackage{hyperref}
\begin{document}

\title{Combined Classical and Quantum Accelerometers For the Next Generation of Satellite Gravity Missions}

\author{Alireza HosseiniArani$^1$, Manuel Schilling$^2$, Benjamin Tennstedt$^1$, Alexey Kupriyanov$^1$, Quentin Beaufils$^3$, Annike Knabe$^1$, Arpetha C. Sreekantaiah$^1$, Franck Pereira dos Santos$^3$, Steffen Schön$^1$, Jürgen Müller$^1$}


\address{$^1$ Institute of Geodesy (IfE), Leibniz University Hannover, Hannover, Germany}
\address{$^2$ Institute for Satellite Geodesy and Inertial Sensing, German Aerospace Center (DLR), Hannover, Germany}
\address{$^3$ LNE-SYRTE, Observatoire de Paris, Université PSL, CNRS, Sorbonne Université, Paris, France}

\ead{hosseiniarani@ife.uni-hannover.de}
\vspace{10pt}
\begin{indented}
\item[]May 2024
\end{indented}

\begin{abstract}

Cold atom interferometry (CAI)-based quantum accelerometers are very promising for future satellite gravity missions thanks to their strength in providing long-term stable and precise measurements of non-gravitational accelerations. However, their limitations due to the low measurement rate and the existence of ambiguities in the raw sensor measurements call for hybridization of the quantum accelerometer (Q-ACC) with a classical one (e.g., electrostatic) with higher bandwidth.  
While previous hybridization studies have so far considered simple noise models for the Q-ACC and neglected the impact of satellite rotation on the phase shift of the accelerometer, we perform here a more advanced hybridization simulation by implementing a comprehensive noise model for the satellite-based quantum accelerometers and considering the full impact of rotation, gravity gradient, and self-gravity on the instrument.
We perform simulation studies for scenarios with different assumptions about quantum and classical sensors and satellite missions. The performance benefits of the hybrid solutions, taking the synergy of both classical and quantum accelerometers into account, will be quantified. We found that implementing a hybrid accelerometer onboard a future gravity mission improves the gravity solution by one to two orders in lower and higher degrees. In particular, the produced global gravity field maps show a drastic reduction in the instrumental contribution to the striping effect after introducing measurements from the hybrid accelerometers.  
 

-

\end{abstract}

%
\vspace{2pc}
%
%
%
\ioptwocol

\section{Introduction}

\subsection{Next generation of satellite gravity missions}\label{gravityMissions}


Satellite gravimetry missions play a crucial role in monitoring the Earth's gravity field and its temporal changes. 
Missions such as GRACE(-FO) have significantly contributed to the understanding of mass variations associated with climate change \citep{Tapley2019,Humphrey2023,Scanlon2023} and gave insights into the Earth's interior processes \citep{Mandea2020, Lecomte2023}.  
The current solutions of the gravity field offered by these missions are limited, particularly at very low degrees such as $C_{20}$ and even extending to $C_{30}$ in times when only one operational accelerometer is available on two satellites,  \citep{Loomis2020}. Consequently, these coefficients are often substituted with values from satellite laser-ranging solutions. Another challenge appears due to the drift in low frequencies of the electrostatic accelerometers employed in the GRACE(-FO) missions, thereby constraining the gravity field solution. In general, the spatial resolution remains confined to areas of a few hundreds of kilometer \cite{Chen2022}  
for a signal amplitude of \qty{10}{\milli\m} in typical monthly gravity field solutions. To meet the demands of the scientific community, future satellite gravity missions must aim for a resolution of $(\qty{200}{\km})^2$ for an amplitude of \qty{10}{\milli\m} equivalent water height or even smaller \citep{Pail2015, Wiese2022}. In this study, we focus on the instrumental contributions to the gravity field mission performance. The other main contributors are addressed elsewhere \citep[\eg][]{Haagmans2020,Shihora2022}. 

\subsection{Quantum accelerometer}\label{introCAI}

Atom interferometry accelerometers implement three laser pulses acting either as beam splitters or mirror pulses \cite{Kasevich1991,Geiger2020}. These laser pulses consist of two counter-propagating laser beams whose frequency difference is tuned in resonance with a two-photon Raman transition between the two hyperfine ground states of an alkali atom, \ce{^{87}Rb} in our case. 
The first pulse of light creates a beam splitter, which places the atom in a quantum superposition of two wave packets of different momenta. 
A second light pulse inverts the momenta of the two wave packets after a time interval $T$. 
Finally, a second beam splitting pulse closes the interferometer after a further time interval $T$.
The atom interferometer phase shift $\Delta\Phi$ is the result of the projection of the acceleration $\vec{a}$ experienced by the atoms along the effective optical wave vector of the laser light $\vkeff=\vec{k}_1-\vec{k}_2$ which is the difference between the optical wave vectors $\vec{k}_i$ of both laser beams.
The leading order of the atom interferometer phase $\Delta\Phi$ is described by
\begin{equation}    \label{eq:CAIphase}
    \Delta\Phi= (2~\keff ~ a)  T^2 + \Phi_{L,i}
\end{equation}
with the acceleration $a$ and $\keff=|\vkeff|$ now expressed in direction of the counter-propagating laser beams. 
$\keff$ is also related to the photon momentum exchange by $\hbar\keff$ where $\hbar$ is the reduced Planck constant. 
A factor $2$ is added because we consider a double diffraction \citep{leveque2009} Q-ACC, as it is anticipated for microgravity (for more details see~\citep{hosseiniarani2024}). 
An arbitrary Raman laser phase $\Phi_L$ can be added to the last light pulse to operate the interferometer, \eg, at mid-fringe as described in~\cite{hosseiniarani2022}. 

According to Eq.~\eqref{eq:CAIphase}, the sensitivity of the Q-ACC can be increased by increasing the interrogation time $T$. In terrestrial applications, $T$ is limited by the length of the free fall distance of the atoms, \eg up to \SI{300}{\milli\s} for a transportable \citep{Freier2016} and up to a couple of seconds for stationary instruments \citep{Asenbaum2020,Schilling2020}. 
As atoms and satellites in space are in free fall, longer separation times $T$ are possible. There, a quantum accelerometer would allow for monitoring the deviation from the free fall trajectory resulting from non-gravitational accelerations acting on the satellite.

The sensitivity limits of an atom interferometer in absolute inertial sensing have been discussed in detail in~\cite{Geiger2020,hosseiniarani2024}. To be brief, with the currently available technology, a space-based CAI-based quantum sensor with some minor improvements is expected to achieve sensitivity in the order of \msqwert{1e-10}. However, ongoing and future advances in atom interferometry in the next 5 to 10 years are expected to reach a sensitivity in the order of \msqwert{1e-11}~\citep{hosseiniarani2024,Zahzam2022,Knabe2022}. In section~\ref{sec:res}, we study the hybridization of quantum sensors with conventional ones based on different assumptions for the sensitivity of the sensors.

\begin{table*}
\centering
\caption{Parameters of the quantum accelerometers considered in different scenarios} \label{tab:advancesTab}
\begin{tabular}{|l|c|c|}
\hline
       \multirow{2}{*}{Parameter}   & Quantum sensor based on & \multirow{2}{*}{Near-future quantum sensor $^b$} \\
          & state-of-the-art technology $^a$ &  \\
\hline
Laser waist  & {\qty{10}{\mm}} & {\qty{20}{\mm}}\\
\hline
Atomic temperature  & \qty{40e-12}{\kelvin} & \qty{10e-12}{\kelvin}\\
\hline
Number of atoms  & \num{5e5} & \num{1e6}\\
\hline
Error on the positioning & \multirow{2}{*}{\qty{3e-5}{\m}} & \multirow{2}{*}{\qty{3e-5}{\m}} \\
 of the atomic cloud $^c$  & &  \\
\hline
Transversal velocity of atoms  & \qty{33e-6}{\m\per\s} &  \qty{20e-6}{\m\per\s} \\
\hline
Error on the measurement   &  \multirow{2}{*}{{\qty{2.2e-7}{\radian\per\s}}} & \multirow{2}{*}{{\qty{6.6e-8}{\radian\per\s}}}\\
of rotation rates & & \\
\hline
\multirow{2}{*}{Rotation compensation} & counter-rotating & counter-rotation of  \\
   & Raman mirror & the quantum sensor  \\
   \hline
Atomic flight time ($2T$)  & \qty{5}{\s} & \qty{10}{\s}    \\ 
\hline
\end{tabular}
\footnotesize{\\ $^a$ In this scenario, minor improvements have been considered for the quantum sensor based on state-of-the-art technology presented in \cite{hosseiniarani2022} to make the quantum sensor compatible with the state-of-the-art classical accelerometers; $^b$ We assume the near future quantum sensor as what can be achieved in the next 5 to 10 years \citep{hosseiniarani2024}; $^c$ for more information about this error please refer to \cite{hosseiniarani2024}.}
\end{table*}

\subsection{Hybridization of the classical and quantum sensors}\label{introHyb}

The benefit of hybridization of cold atom and electrostatic accelerometers for gravity field missions in a simplified scenario without rotational effects has already been shown \citep[\,  e.g.][]{Abrykosov2019}.
These studies typically generate noise-only time series for the two accelerometers and combine them, \eg, by filtering.
The hybrid accelerometer noise, converted to ranging accelerations, is added to the ranging observations before gravity field recovery.
Our method combines the (noisy) measurements of the two accelerometers while simultaneously using the measurements of the electrostatic accelerometer to solve the phase ambiguities of the cold atom interferometer.
The method introduced here can potentially be used in real-time scenarios for data generated by future hybrid sensors and has already been demonstrated for application in inertial navigation using a navigation-grade IMU to emulate a CAI, \citep{Weddig2021}.

In our previous study \citep{hosseiniarani2022}, we have shown the benefit of such hybridization in a simplified case for a GRACE-like mission. Here, we realize a more advanced version of the hybridization in which we use a comprehensive noise model for the quantum accelerometers and model the full impact of rotation on the Q-ACC and the Kalman filter.

Three scenarios were studied: first, the hybridization of a Q-ACC based on state-of-the-art technology with a few minor improvements with an electrostatic accelerometer similar to the one on-board of the GRACE-FO satellites. In the second scenario, we study the hybridization of a highly improved Q-ACC (what we expect to achieve in the next 5 to 10 years) with another proposed highly accurate electrostatic accelerometer model. For the first two scenarios, we assumed the satellites to be at an altitude of around \qty{480}{\kilo\m}. In the third scenario, we assume the same sensors as the first scenario, but the satellites were assumed to be in an orbit of \qty{300}{\kilo\m} altitude. The latter would show the efficiency of the filter in the presence of high-amplitude non-gravitational signals.

In section~\ref{sec:modeling}, we describe our modeling for the satellite gravity missions and the classical and quantum accelerometers onboard. We then discuss the modelling of the rotation, gravity gradient, and self-gravity effects. Afterwards, we introduce the Kalman filter approach we use to hybridize the accelerometers' measurements and discuss the simulation of the recovery of the gravity field. In section~\ref{sec:res}, we discuss the results of the hybridization in three different scenarios, and we compare the achievable accuracy in the gravity field recovery by introducing the hybrid accelerometer concepts onboard future satellite gravity missions.

\section{Modeling}  \label{sec:modeling}

\subsection{Modeling a GRACE-like gravity mission}
In this study, we consider a GRACE-like satellite pair in a circular polar orbit around the Earth with an altitude of \qty{480}{\kilo\m}. The simulation is implemented in the MATLAB/Simulink-based eXtended High-Performance satellite dynamics Simulator \citep[XHPS, ][]{Woeske2019} developed by ZARM/DLR. XHPS calculates the orbits of a GRACE-FO mission scenario under consideration of the Earth’s gravity field ‘EGM2008’, non-gravitational forces (atmospheric drag, solar radiation pressure, Earth albedo and thermal radiation pressure) and the GRACE satellite geometry. To consider the effect of non-gravitational forces on the spacecraft, we use a detailed surface model of the satellite body included in XHPS.

\subsection{Classical and quantum sensor models}

\begin{figure}

\includegraphics[width=\columnwidth]{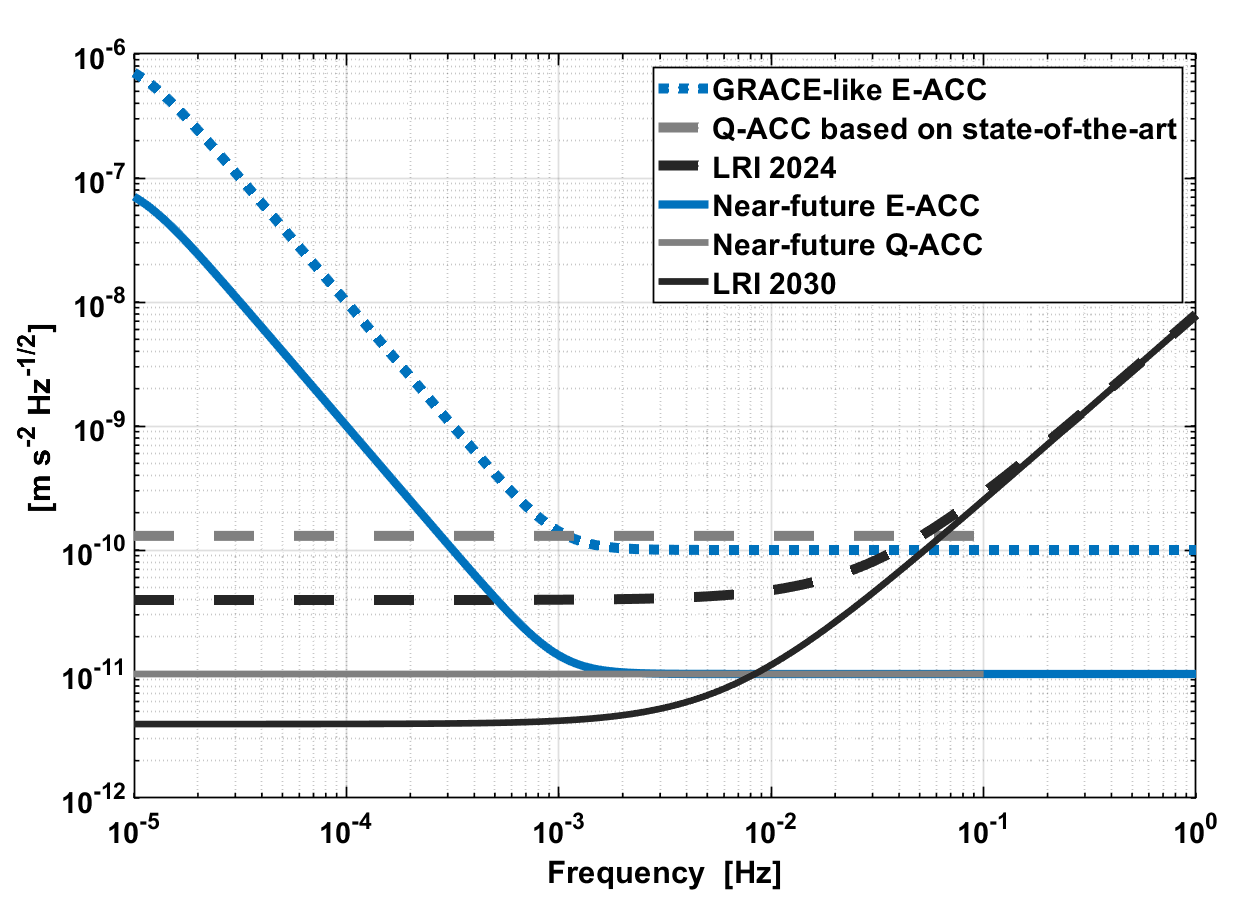}
\caption{Amplitude spectral densities in range accelerations in
the along-track direction for a GRACE-like satellite gravity mission using an electrostatic (classical) and CAI (quantum) accelerometers for the state-of-the-art and near-future scenarios \citep{Flury2008,Christophe2015,hosseiniarani2024}; together with laser ranging
systems \citep{Abich2019,KUPRIYANOV2024}.}
\label{fig:EACAIASD}
\end{figure}

An electrostatic accelerometer (E-ACC) is a classical sensor that measures accelerations. If the E-ACC is placed at the satellite's center of mass, it only measures the non-gravitational accelerations acting on the satellite in three orthogonal directions (along-track, cross-track, and radial). This is because the satellite is free-falling in the Earth's gravity field. 
A classical or conventional E-ACC usually performs best in a frequency range higher than \SI{e-3}{\hertz}, while at lower frequencies, the measurements suffer from large noise. 
The low-frequency noise  (see figure~\ref{fig:EACAIASD}) causes a bias in the measured accelerations. The accelerometer measurements, therefore, can be written as: 
\begin{equation}
A_{ACC} = B + S \cdot A_{non\text{-}grav.} + N
\label{eq:eacc_model}
\end{equation}
where $B$ is the accelerometer bias, $S$ is the accelerometer scaling factor and $N$ stands for the random noise. In this study, we ignore the scaling factor and focus on the determination of the E-ACC bias. The sensor model of an E-ACC, based on the GRACE ACC sensitive axis \cite{Christophe2015} with a noise level of \msqwert{e-10} in frequencies above \SI{e-3}{\hertz}, is implemented in XHPS. The sampling rate of the E-ACC is \SI{10}{\hertz}. 

Although better performing E-ACC have already been in space \citep[\eg GOCE; ][]{Marque2010} or are published as recent developments \citep{Christophe2018, Zahzam2022}, the GRACE-FO type accelerometer was chosen to later compare with GRACE-FO gravity field solutions. This is also a worst-case scenario, i.e., if this method works successfully with this accelerometer, an accelerometer with lower noise at low frequencies is not a priority when enabling a hybrid accelerometer. It is worth mentioning that this type of accelerometer is now again foreseen for NASA's GRACE-C mission. 
Later during this study, we will also use a more accurate model of E-ACC with a noise level of \msqwert{e-11} above \SI{e-3}{\hertz} and test the performance of its hybridization with the Q-ACC model.




In a satellite setting, the Q-ACC can be used to measure the non-gravitational accelerations acting on the satellite. A three-axis instrument could also be implemented for geodesy applications, but in this paper, we focus on a single-axis quantum accelerometer oriented in the along-track (X-) direction. Eq.~\eqref{eq:CAIphase} describes a single-axis accelerometer phase shift in the absence of rotation of the satellite. 
Because of the change of the non-gravitational acceleration during the Q-ACC interrogation time, we integrate this equation by considering the sensitivity function of Q-ACC as described in \cite{Knabe2022}. 




To simulate the error budget of the Q-ACC, we implement a detailed noise model including all the major noise sources that affect the measured phase shift: laser frequency, wavefront aberrations, and detection noises. The details of this error-source model are presented in~\cite{hosseiniarani2024}. 

The sensitivity of space-based quantum accelerometers based on state-of-the-art technology is on the order of \msqwert{5e-10}. To have comparable accuracy with a GRACE-like accelerometer, we consider a slightly optimized version of the state-of-the-art scenario discussed in~\cite{hosseiniarani2024} (see table~\ref{tab:advancesTab}), and we study its hybridization with a GRACE-like electrostatic accelerometer. 

In another scenario, we consider a highly improved Q-ACC, which we expect to be achieved in the next 5 to 10 years, equivalent to the near-future scenario given in~\cite{hosseiniarani2024}, and we study its hybridization of an improved electrostatic accelerometer. Figure~\ref{fig:EACAIASD} shows the expected performance of the quantum and electrostatic sensors for the two scenarios mentioned. 

Note that in all our scenarios, we assume a three-axis E-ACC combined with a one-axis Q-ACC with its sensitivity axis aligned to the along-track direction of the satellite. The motivation for keeping the sensitivity axis along-track is that the non-gravitational acceleration in flight path is the most critical component for a GRACE-like mission, i.e. the acceleration of the satellite along this direction directly affects the determination of the gravity field. 

%

\subsection{Effect of satellite rotation} \label{sec:rotModel}

When the quantum sensor is in a rotating body frame, e.g. a nadir-pointing satellite, based on equation \eqref{eq:CAIphasePlusRot}, there will be an additional phase shift caused by the Coriolis and centrifugal accelerations. 
The largest contribution of this rotation to the phase shift of the atom interferometer arises from the Coriolis acceleration induced by the atomic velocity in the radial direction \citep{Leveque2021,Lan2012}. Also, centrifugal and Euler accelerations lead to additional terms. 
In our previous study of the hybrid accelerometer \citep{hosseiniarani2022}, we assumed to have a perfect knowledge of the rotational accelerations, and therefore, only the measurements of linear accelerations by an E-ACC were considered as the measurements of conventional inertial measurement unit. In this study, we consider the whole rotational effect. To do so, we first model the angular dynamics of the satellite and an attitude control system for a GRACE-like satellite. Then, we model the measurement of gyroscopes by adding white noise to the true angular velocities. Finally, we model the effect of the rotation on the measurements of the Q-ACC. 
If the rotation of the satellite is not properly compensated, the solution will degrade considerably. Therefore, several approaches are proposed to physically compensate the effect of the rotation \citep[see ][]{Zahzam2022, Beaufils2023}. 

\begin{figure}

\includegraphics[width=\columnwidth]{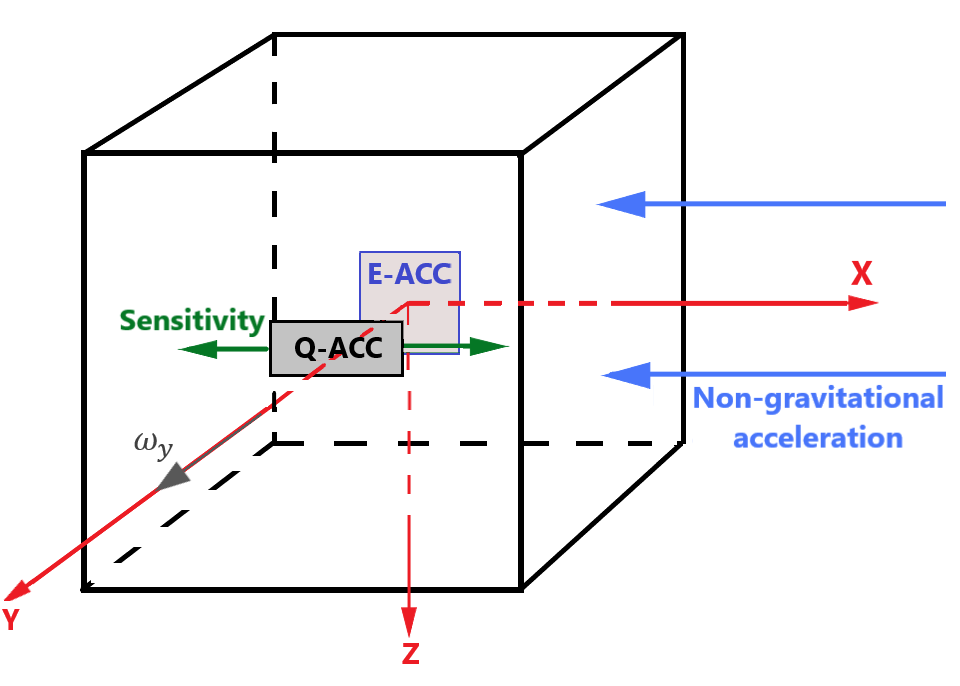}
\caption{Positioning of the quantum accelerometer inside the satellite frame considering that the E-ACC is in the center of mass. The Q-ACC accelerometer is placed on the cross-track axis of the satellite. The sensitivity axis of the Q-ACC is parallel to the along-track axis of the satellite.}%
\label{fig:CAIpos3}%
\end{figure}

The positioning of the classical and quantum accelerometers inside the satellite frame, as well as the rotational compensation method that is used against the satellite's rotation around the cross-track axis, have a considerable impact on the accuracy of the Q-ACC. Comparing these rotation compensation methods is out of the scope of this study. Here, we consider only one of the optimal situations in which the classical sensor is placed at the center of mass of the satellite and the quantum sensor is placed beside it on the cross-track axis of the satellite (see figure~\ref{fig:CAIpos3}). Note that the sensitivity axis of the quantum sensor is still in the along-track direction, so we only measure the accelerations in this direction which is the main direction of interest for GRACE-like missions. This configuration minimizes the impact of centrifugal accelerations by minimizing the lever arm. 
The two main rotation compensation methods use active counter-rotating Raman mirrors and counter-rotating the whole quantum sensor against the satellite rotation around the cross-track axis. 
The implementation of these methods in our simulations is described in \cite{hosseiniarani2024}.
Here, we use the former method for our first scenario and the latter for our second scenario. We then study the hybridization in both scenarios. The phase shift due to the satellite rotation can be calculated 
based on \cite{Beaufils2023}:

  \begin{equation}    \label{eq:CAIphasePlusRot}
\begin{split}
    & \Delta\Phi= 2 \keff T^2 [a_x + 2 v_{z0} (\Omega_y + \Omega_M) - x_0 {\Omega_y}^2 \\
    & + (x_0 - x_M)({\Omega_M}^2 + {(\Omega_M - \Omega_I)}^2)]
\end{split}
\end{equation}

where $x_0$ is the initial distance of atoms to the satellite center of mass, $x_M$ is the distance of the center of rotation of the mirror to the satellite center of mass, $\Omega_y$ is the angular velocity of the satellite around the cross-track axis with respect to the inertial frame, $\Omega_M$ and $\Omega_I$ are the angular velocities of the mirror and incoming laser beam, with respect to the satellite body-fixed frame, and $v_{z0}$ is the initial velocity of atoms in the radial direction of the satellite body frame. 
In the case of rotation compensation using an active counter-rotating mirror, a high-performance onboard gyroscope can be used to measure the satellite rotation at each instant in time and cancel its contribution to the phase shift by an active Raman mirror rotating against the rotation rate of the satellite \citep{Lan2012,Migliaccio2019}. In this scenario, the incoming laser is assumed to be fixed in the body frame of the satellite and has a rotation rate identical to that of the satellite ($\Omega_I = 0$ and $\Omega_I = 0$). We also assume that the center of rotation of the mirror is the same as the center of mass of the satellite ($x_M = 0$). Therefore, the equation~\ref{eq:CAIphasePlusRot} can be simplified as follows

\begin{eqnarray}    \label{eq:mirRot2}
  \Delta\Phi= 2 \keff T^2 [a_x + x_0 {\Omega_y}^2].
\end{eqnarray}

The term $x_0 {\Omega_y}^2$ is the remaining term after compensation of rotation by a counter-rotating Raman mirror. In the case of rotation compensation by counter-rotating the entire sensor, this remaining term will also be removed from the phase shift equation. However, technical difficulties might arise in the realization of this method. In both scenarios, the effectiveness of rotation compensation will be highly dependent on our knowledge of the rotation rates. The details of these formulations are discussed in ~\cite{hosseiniarani2024}.





\begin{table*}
\centering
\caption{Different signal contributions to the phase shift} \label{tab:phaseShift}
\begin{tabular}{|l|c|c|}
\hline
 Contributions to the phase shift &  size (rad) & fractional size \\
 \hline 
 Non-gravitational acceleration &  \num{19.97} & 1 \\
 Uncompensated rotation &   \num{8.6e-3} & \num{4.3e-4}\\
 Self-gravity &   \num{4.8e-5} & \num{2.4e-6}\\
 Gravity gradient  ($V_{xx}$) &   \num{3.1e-06}~$^a$ & \num{1.5e-7}\\

\hline 

\end{tabular}
\footnotesize{\\ $^a$ This number is calculated for the interrogation time ($2T$) of $\qty{10}{\s}$; if the interrogation time is considered to be \qty{5}{\s}, then this phase shift contribution would be \qty{3.87e-7}{\radian}}
\end{table*}

\subsection{Gravity gradient and self-gravity}

As shown in figure~\ref{fig:CAIpos3}, the quantum sensor is assumed to be displaced from the centre of mass of the satellite, which leads to an additional phase shift due to the gravity gradient. However, since the sensitivity axis of the quantum sensor is assumed to be in the along-track direction, only the gravity gradient $g_{xx}$ in the along-track direction results in an additional phase shift. Our simulations show this effect has a small value and, therefore, can be neglected (see table~\ref{tab:phaseShift}). 

Self-gravity is the gravitational pull of the spacecraft's mass on the free-falling atoms. Depending on the position of the quantum sensor and the direction of its sensitivity axis, the self-gravity effect can also result in an additional phase shift. Our simulations show that if the quantum sensor is placed on the along-track axis, a \qty{20}{cm} displacement can cause a considerable phase shift in the order of \qty{0.27}{\radian}. However, by having knowledge of the satellite mass and the displacement, this phase shift can be calculated and removed from the measurements. Moreover, if the quantum sensor is placed on the cross-track axis of the satellite while its sensitivity axis is still in the along-track direction (similar to the situation in figure~\ref{fig:CAIpos3}), the self-gravity effect would be minimal. Table~\ref{tab:phaseShift} shows the self-gravity phase shift in this scenario, which is around two orders of magnitude smaller than the total rms phase noise of the sensor and can be neglected.

In addition to the above-mentioned signals, there are higher-order contributions that couple these inertial forces, particularly with gravity gradients. However, because of the smaller contributions of these additional terms and since the sensitivity axis of the Q-ACC is aligned in the along-track direction of the satellite and then the gravity gradient in the along-track direction is considerably smaller than in the radial direction, the additional phase shift gets relatively small values. Table~\ref{tab:phaseShift} compares the contributions of the non-gravitational accelerations, rotation, gravity gradient and self-gravity effects on the total phase shift of quantum accelerometers. Given the comparatively small values of the phase shift due to the gravity gradient and self-gravity, we ignore these terms in our modeling and take into account only the rotation terms.


\subsection{Extended Kalman Filter} \label{sec:filter}

Our extended Kalman Filter (EKF) seeks to combine the measurements of the two accelerometers while simultaneously using the measurements of the electrostatic accelerometer to solve phase ambiguities of the Q-ACC.
In essence, the phase ambiguity solution is realized by a prediction of the Q-ACC observation with the data of the E-ACC and gyroscope, using a model according to equation~\eqref{eq:CAIphasePlusRot}, in which the acceleration $a_x$ is assessed by the E-ACC model from Eq.~\eqref{eq:eacc_model}.

The predicted observation of the quantum accelerometer $h_i$ at a discrete time indicated by index $i$ is produced by the predicted phase shift $\Delta \Phi_i$. It reads
\begin{equation}
h_i(\Delta \Phi_i) = A\cos{(\Delta \Phi_i + \phi_{L,i})} + \nu.
\end{equation}
The fringe parameters $A$ and $\nu$ are each assumed to have a value of $0.5$. It is further anticipated that the laser phase $\phi_{L,i}$ is steered, such that the sum of $\Delta \Phi_i + \phi_{L,i}$ equals $\pi/2$. 
An estimate $\hat{B}_i$ of the E-ACC is then obtained based on the difference of this prediction and the actual Q-ACC observation $p_{CAI,i}$ via 
\begin{equation}
\hat{B_i} = -K_i(p_{CAI,i} - h_i(\Delta \Phi_i)).
\end{equation}
This innovation $p_{CAI,i} - h_i(\Delta \Phi_i)$ is weighted by the Kalman gain $K_i$, 
\begin{equation}
K_i = \frac{P_i^- H_i}{R_i + H_i^2 P_i^-},
\label{eq:KGain}
\end{equation}
which itself is related to the \textit{predicted} state variance $P^-_i$, the observation variance $R_i$ and the observation sensitivity $H_i$, 
\begin{equation}
H_i=\frac{\partial h_i}{\partial B} = 2k_\text{eff}T^2,
\end{equation}
of the model to the bias. 
The state variance $P^-_i$ is predicted before the evaluation of Eq.~\eqref{eq:KGain} via
\begin{equation}
P^-_i = P^+_{i-1} + Q_i.
\end{equation}
Here, $Q_i$ is the process noise which corresponds to the uncertainty of the E-ACC. $P^+_{i-1}$ is the \textit{filtered} variance of the prior update, which is defined by
\begin{equation}
P^+_i = (1- K_i H_i)^2 P^-_i + K_i^2 R_i.
\end{equation}

This method was already applied in \cite{hosseiniarani2022}. Since equation \eqref{eq:CAIphasePlusRot} does not yield any additional terms including $a_x$, and thus no additional sensitivities to $B$, no major adaptions of the settings are necessary. However, in the presence of rotation, the determination of the Q-ACC phase ambiguity could be affected in some cases.

The effect of satellite rotation on the measurements of the Q-ACC is discussed in section~\ref{sec:rotModel}. 
In our Kalman filtering strategy, we use the phase shift equivalent to the acceleration measured by the E-ACC to determine the phase ambiguity of the quantum sensor. Since the rotation can cause an additional phase shift, and because of the noisy nature of the measurements of rotation rates by gyroscopes, there will be an additional error in the determination of the quantum sensor phase ambiguity with respect to the situation in which no rotational noise is assumed. 
For additional insight into the general EKF framework, the reader is referred to \cite{tennstedt2023atom}.





\subsection{Recovery of the gravity field} \label{sec:gravRecSec}

We investigate the effect of combined accelerometers in the recovery of the gravity field. For this part, a closed-loop simulation procedure was used \citep{KUPRIYANOV2024} in which, at first, the orbital simulations were carried out in XHPS software \citep{Woske.2016, Woske.2018}.
Then, computed satellite positions were utilized in the gravity field recovery software that follows the procedure introduced by \cite{Wu2016}. And finally, retrieved gravity field models were compared with the reference one (EIGEN-6C4) by calculating residuals and plotting them in the spatial (as global maps) and spectral (as degree RMS graphs) domains.

\begin{figure}[!tbp]%
\centering
\includegraphics[width=\columnwidth]{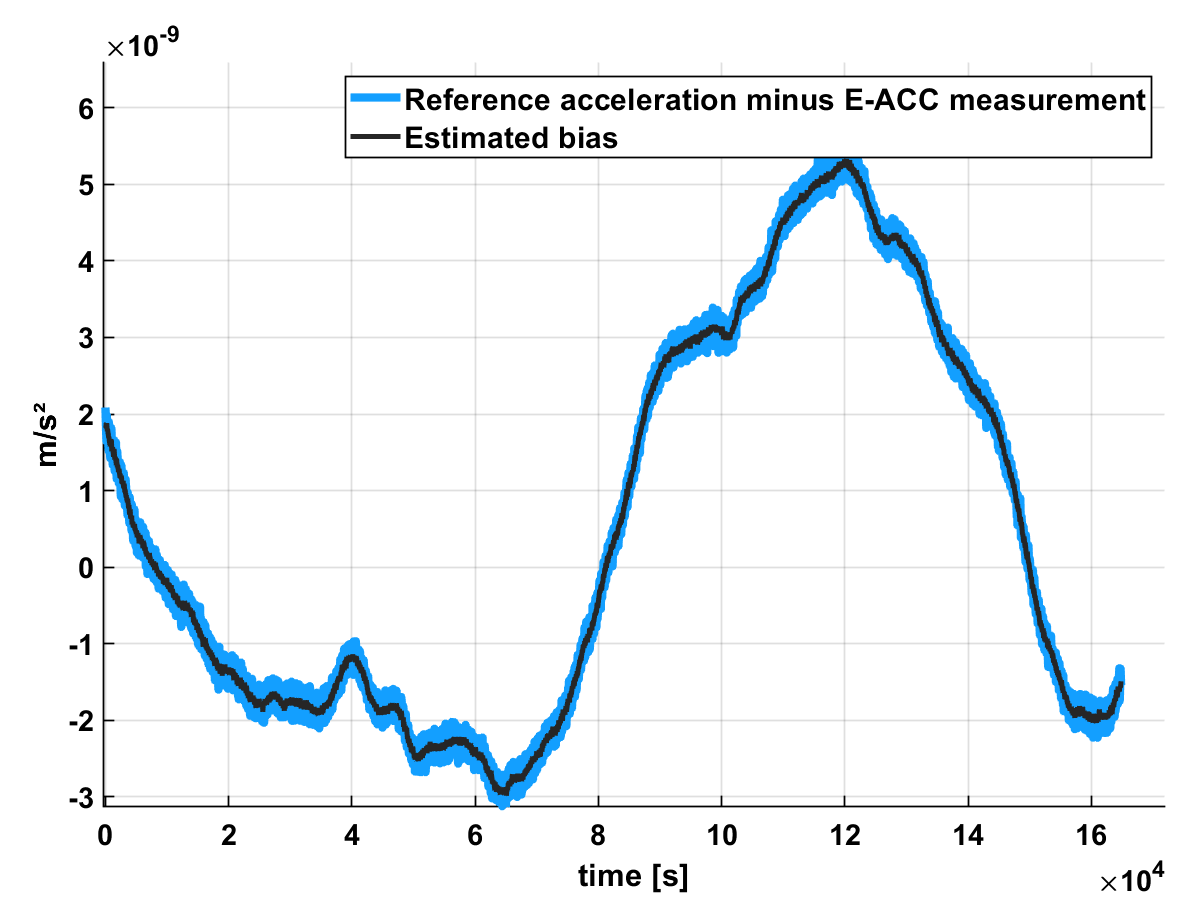}
\caption{Bias estimation, using the hybrid accelerometer concept for the state-of-the-art scenario; Blue: The difference between the true non-gravitational acceleration signal and the measurements of E-ACC; Black: Estimated bias using extended Kalman filtering}%
\label{subfig:estBias1}%
\end{figure}

\begin{figure*}
\includegraphics[width=\textwidth]{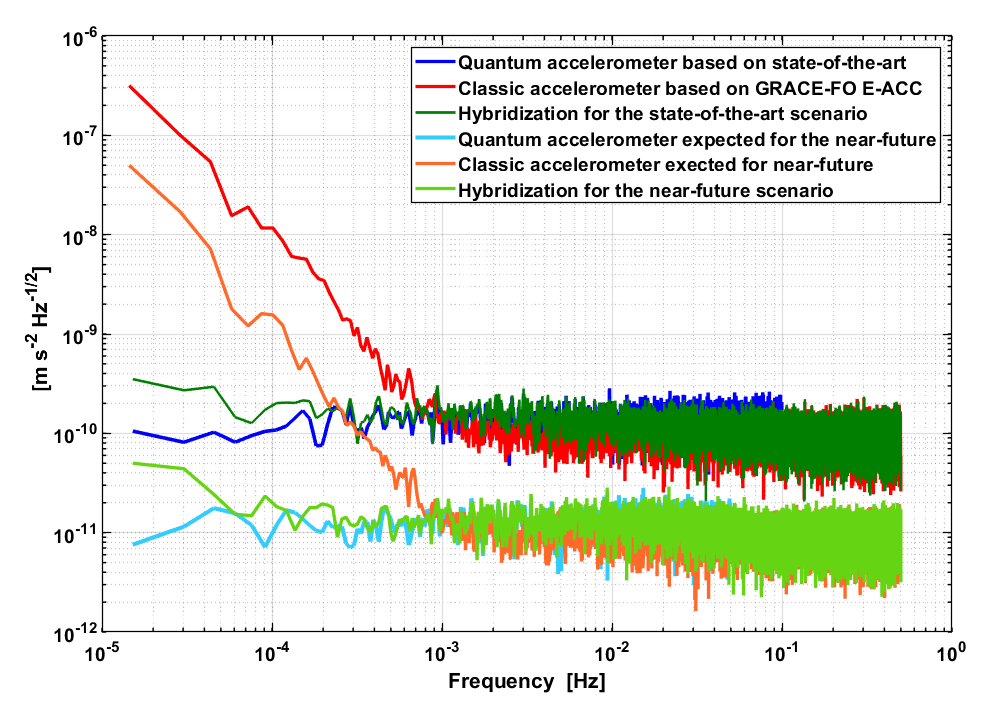}
\caption{Hybridization of classical and quantum accelerometers for the state-of-the-art and near-future scenarios; spectral representation of the solutions in terms of amplitude spectral density
for quantum accelerometer (dark and light blue), classical accelerometers (red and orange), and the Kalman-filter-based hybridization (dark and light green) }%
\label{fig:filterASD}%
\end{figure*}

\section{Results} \label{sec:res}

\subsection{Performance of the hybrid accelerometer for the state-of-the-art scenario} \label{sec:hyb1}

We investigate the efficiency of the Kalman filter in three different scenarios. In the first scenario, we consider an optimized version of the state-of-the-art Q-ACC based on \cite{hosseiniarani2024} and the characteristics described in table \ref{tab:advancesTab} in a hybrid combination with a state-of-the-art E-ACC and similar to the GRACE-FO accelerometer as discussed in section \ref{sec:modeling}. 
Assumptions for the noise of the accelerometers are shown in figure~\ref{fig:EACAIASD}. 


In section~\ref{sec:filter}, we described how we estimate the bias of the E-ACC  $(\hat{B})$ based on the difference of the prediction and the actual CAI observation. Figure~\ref{subfig:estBias1} shows the estimated bias as an output of the EKF compared to the true E-ACC bias (true non-gravitational signal minus the measurements of the E-ACC).

\begin{figure}[!tbp]%
\includegraphics[width=\columnwidth]{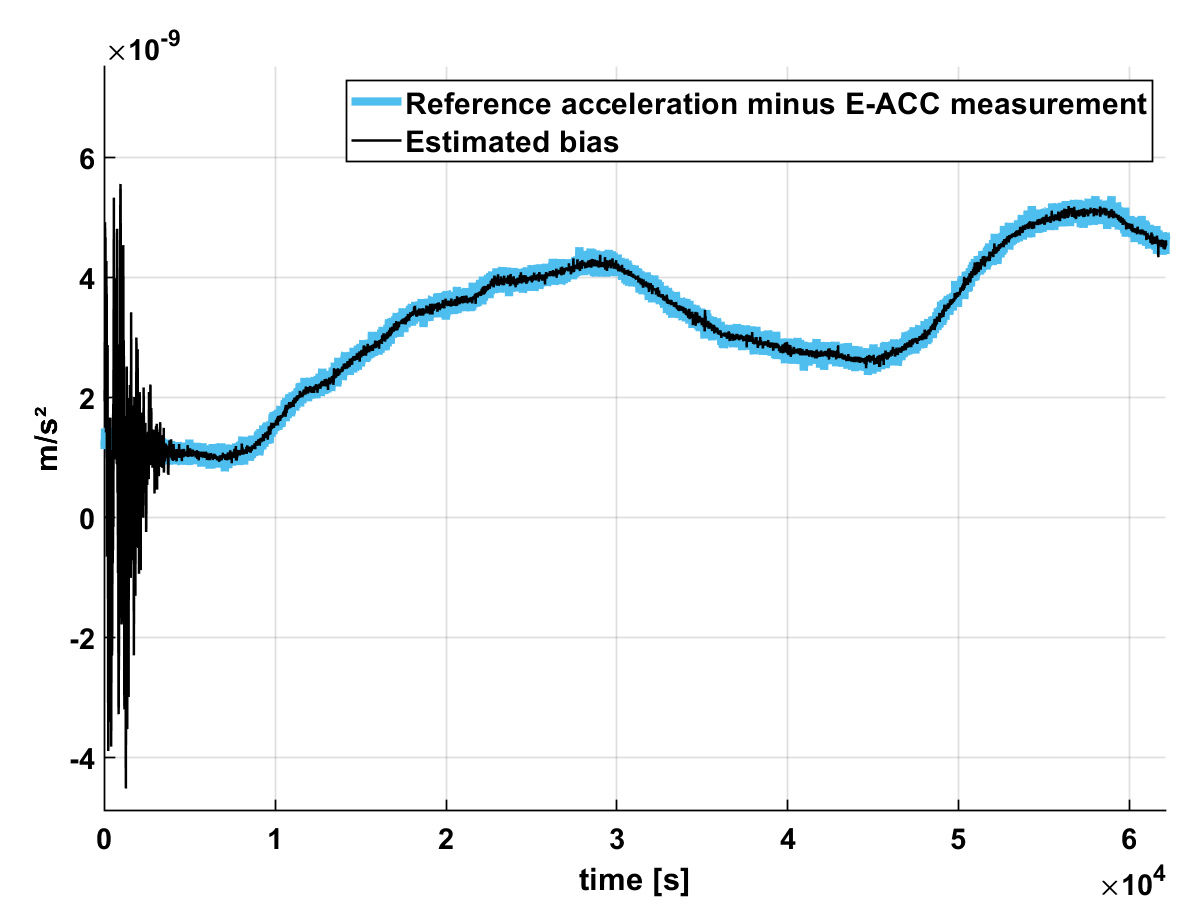}
\caption{Bias estimation, using the hybrid accelerometer concept with the satellite pair being at the altitude of \qty{300}{\km}; Blue: The difference between the true non-gravitational acceleration signal and the measurements of the E-ACC; Black: Estimated bias using extended Kalman filtering}%
\label{fig:estBias4}%
\end{figure}

\subsection{Hybridization for the near-future scenario } \label{sec:hyb2}

New E-ACCs have been demonstrated and tested that can achieve considerably better sensitivity at higher frequencies. 
At the same time, with the ongoing advances in atom interferometry, improved quantum accelerometers are also expected to be realized. We have summarised the impact of key parameters, such as atomic number and temperature, laser system, rotation sensing and mitigation schemes, and their potential for improvement on instrument performance in \citep[][]{hosseiniarani2024}. Therefore, in our second scenario, we consider one order of magnitude improvement in the noise of the electrostatic accelerometer. To have a meaningful combination, we assume a near-future Q-ACC as described in~\cite{hosseiniarani2024} reaching a sensitivity of the level of \msqwert{1e-11} (see figure~\ref{fig:EACAIASD} for the accelerometer noise assumptions). 


Figure~\ref{fig:filterASD} compares the amplitude spectral densities of the quantum and electrostatic accelerometers with the filter solution for both above-mentioned scenarios, where the modeled non-gravitational accelerations are removed. The filter output has gained the stability of Q-ACC measurements at lower frequencies and benefited from the capabilities of the E-ACC at higher frequencies.

\cite{hosseiniarani2024} displays how the rotational phase shift can considerably reduce the interferometer contrast if not properly compensated. It also discusses the impact of different rotation compensation methods on the observation of the quantum sensor.
In section~\ref{sec:filter}, we discussed the additional error that could affect the determination of the quantum sensor ambiguity and the convergence of the Kalman filter. This error is due to the noise on the measurements of the rotation rate compared to the situation where we assume to have perfect knowledge of the rotation. If a proper rotation compensation method is used \citep[based on ][]{hosseiniarani2024}, and the rotational rates are known with the expected accuracy (see table~\ref{tab:advancesTab}), this additional error will not prevent from properly removing the phase ambiguity, nor affect the convergence of the filter.






\subsection{Hybridization at lower altitudes} \label{sec:hyb-alt}
In our third scenario, we consider the same accelerometers that were discussed in our first scenario. However, this time, we assume to have GRACE-like satellites at a lower altitude of \qty{300}{\km}. We want to see if the increase in the amplitude of the non-gravitational and rotational accelerations would have any impact on the filter output.

Figure~\ref{fig:estBias4} compares the recovered bias with the true simulated bias of the electrostatic accelerometer for our third scenario. In this scenario, the filter experiences initial high-amplitude fluctuations in the recovery of the bias, but apart from that, no other meaningful difference in the behaviour of the filter is observed. 
These initial fluctuations are a direct result of the initial wobbling of the satellite in our simulations. They can destroy the filter solution and result in an incorrect bias recovery. However, starting the filter with some delay after the start of the simulation avoids the fluctuations and leads to a filter noise similar to what is shown in dark green in figure~\ref{fig:filterASD}. 



\subsection{Performance in the recovery of the gravity field}

The non-gravitational accelerations in the along-track direction of a GRACE-like mission directly affect the determination of the gravity field. Therefore, by adding a Q-ACC in the along-track direction, we aim for a better determination of the gravity field. Here, we investigate the performance of the combined classical and quantum accelerometers for recovering the Earth's gravity field.

\begin{figure}
  \centering
  \includegraphics[width=\columnwidth]{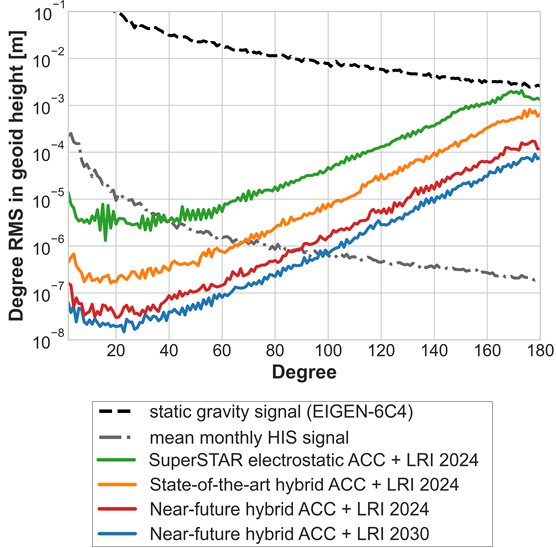}
  \caption{Degree RMS of the spherical harmonic coefficient differences (true errors) plotted in geoid height [m]}
  \label{fig:gravRec}
\end{figure}

\begin{figure*}
    \begin{subcaptiongroup}
\begin{subfigure}{.5\textwidth}
    \centering
    \includegraphics[width=1\linewidth]{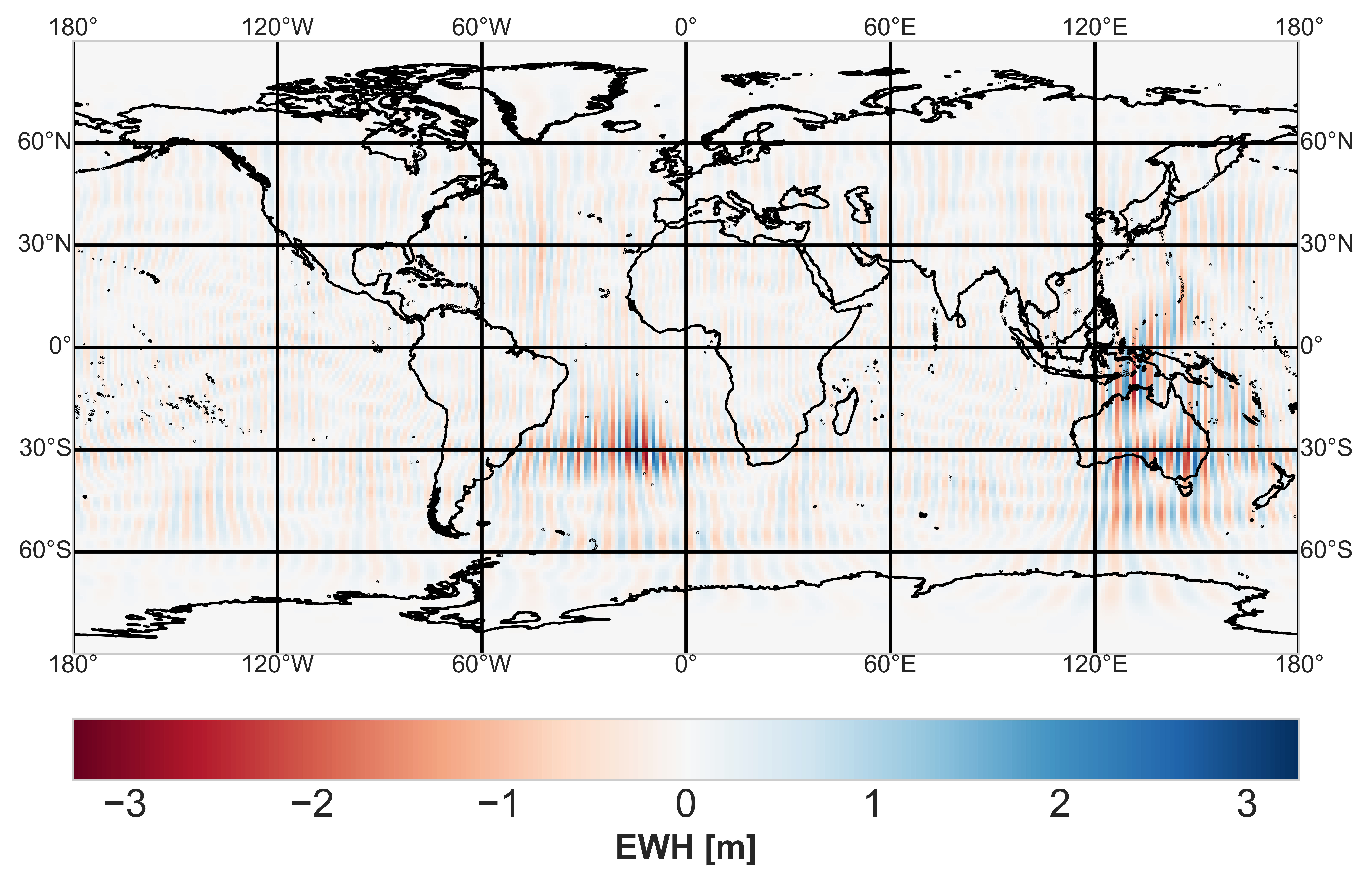}  
\end{subfigure}
\begin{subfigure}{.5\textwidth}
    \centering
    \includegraphics[width=1\linewidth]{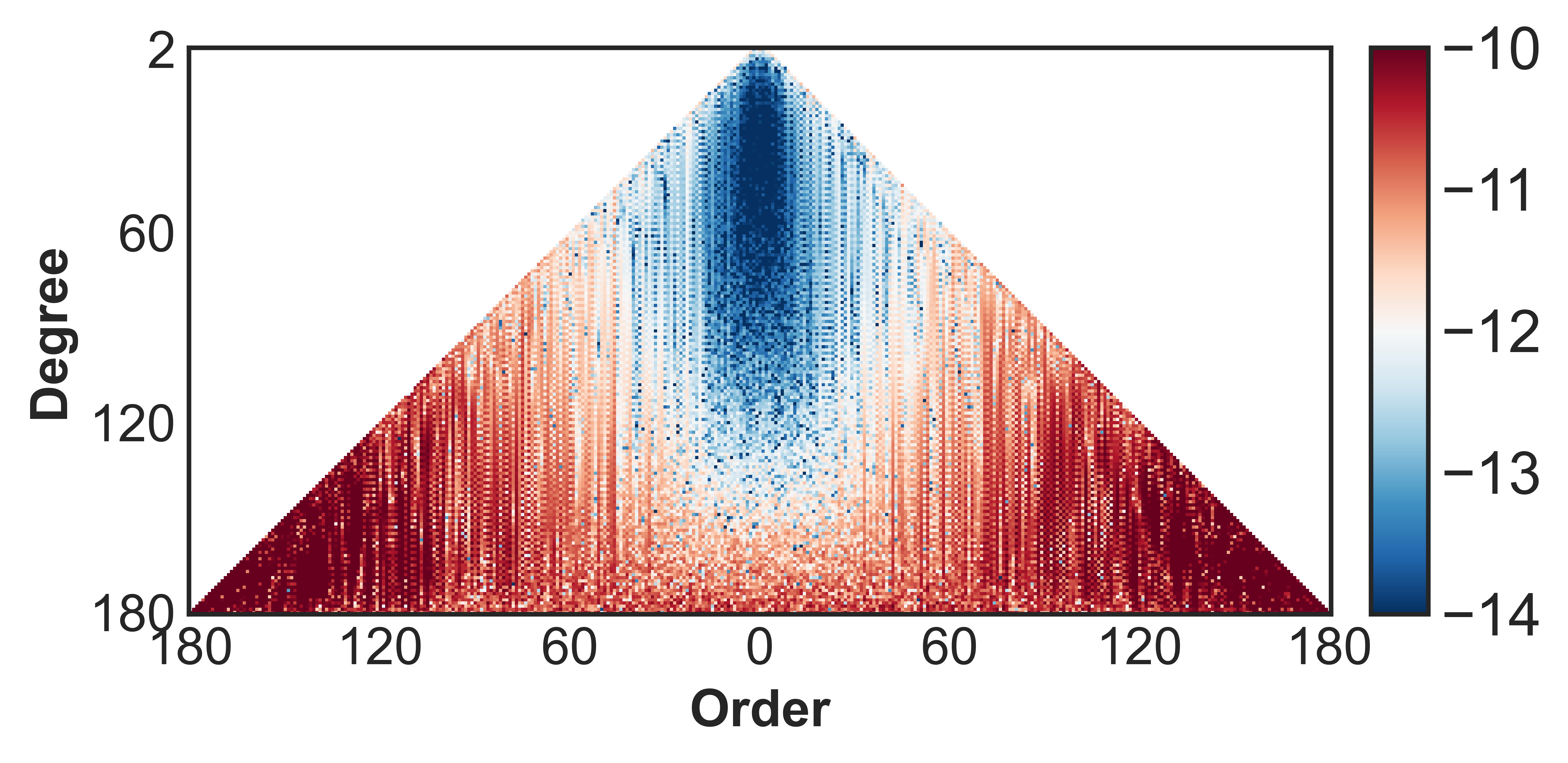}  
\end{subfigure}
\subcaption{SuperSTAR E-ACC $+$ KBR}
\label{subfig:GRV_GRACE}
\begin{subfigure}{.5\textwidth}
    \centering
    \includegraphics[width=1\linewidth]{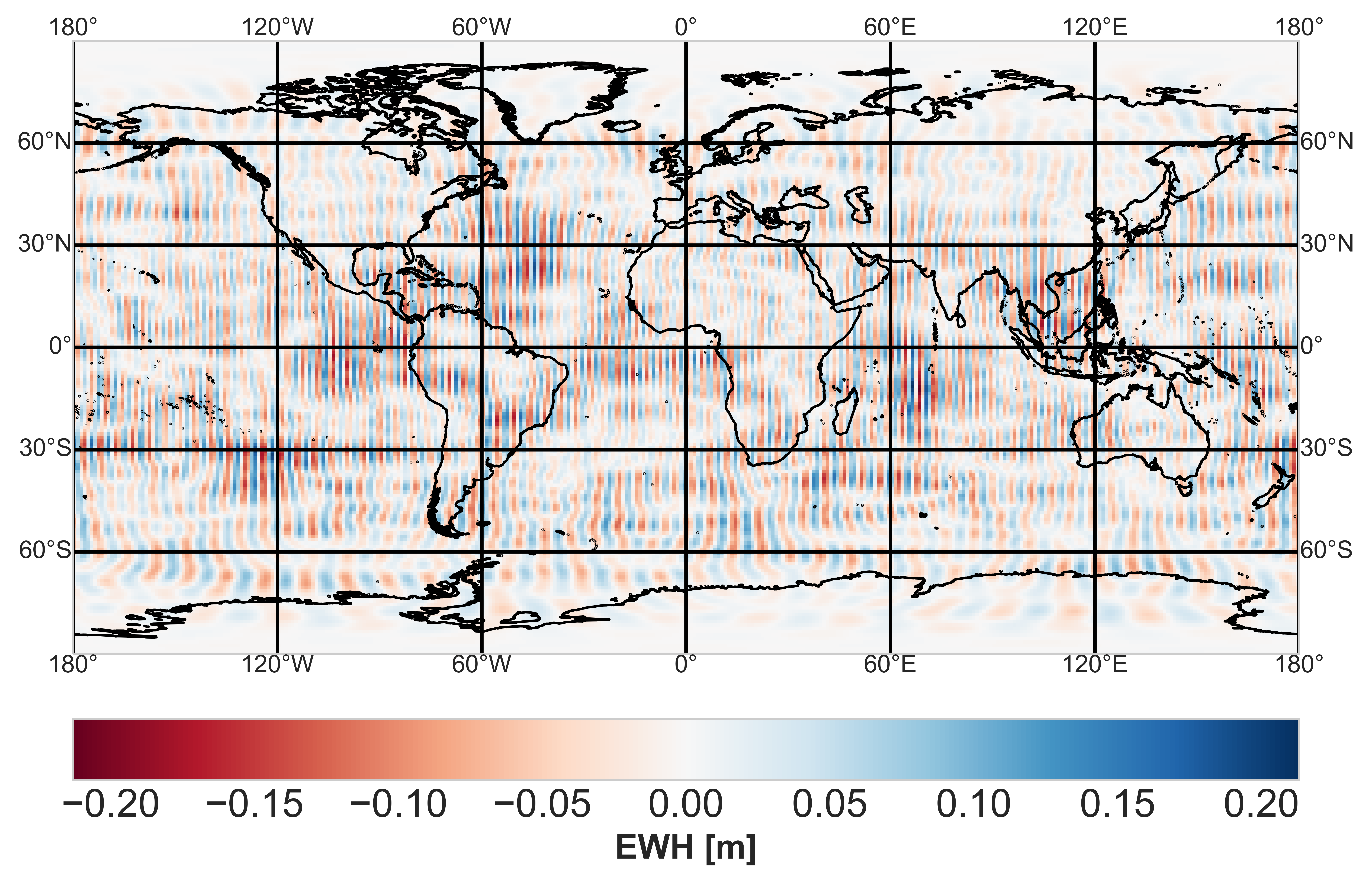}  
\end{subfigure}
\begin{subfigure}{.5\textwidth}
    \centering
    \includegraphics[width=1\linewidth]{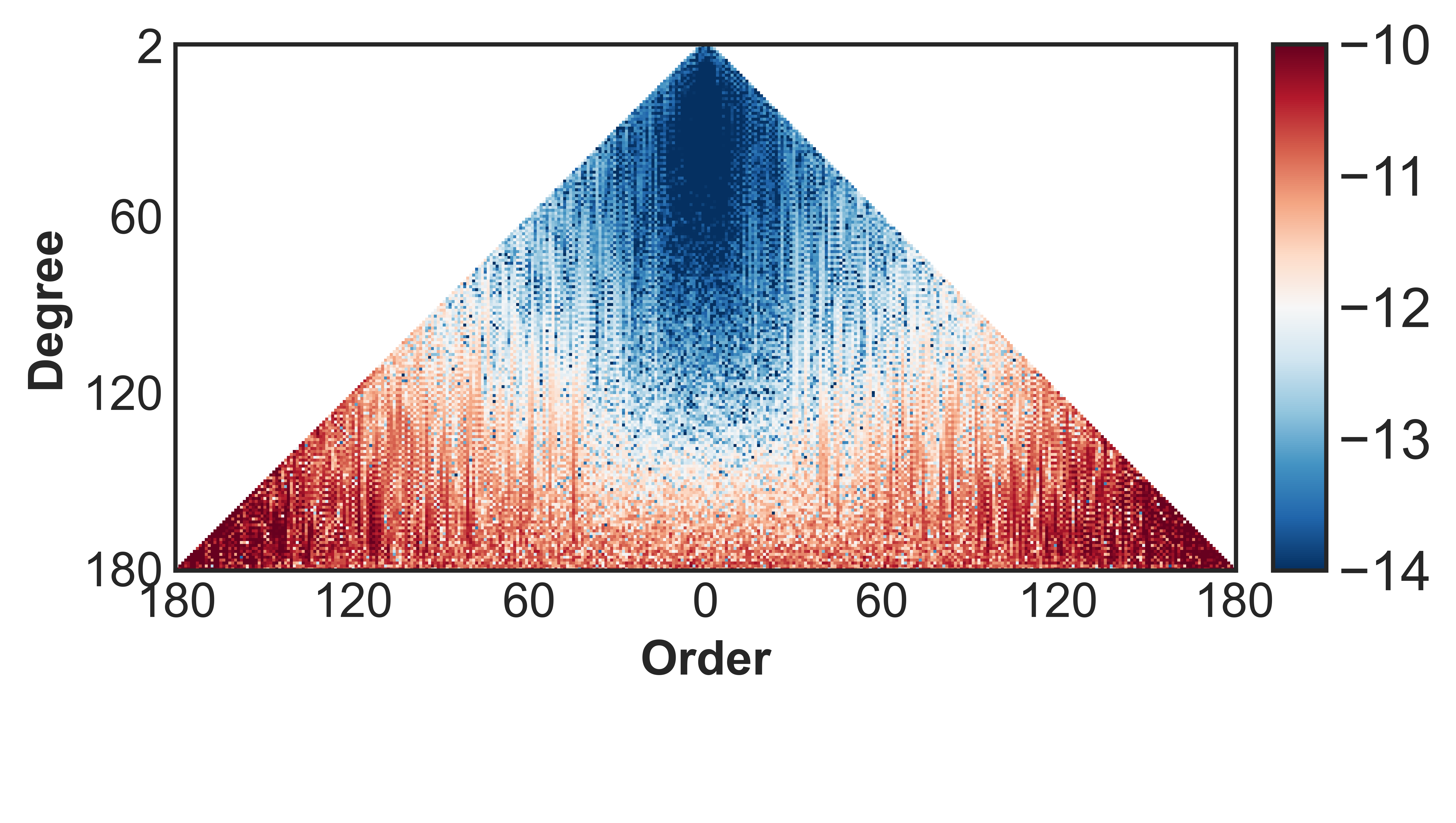}  
\end{subfigure}
\subcaption{State-of-the-art Hybrid ACC $+$ LRI 2024}
\label{subfig:GRV_stateArt}

\end{subcaptiongroup}
\caption{Comparison of the spatial distribution of the residuals plotted on the global maps of the recovered gravity field w.r.t. reference EIGEN-6C4 (without post-processing and filtering) from the GRACE mission (a) and the state-of-the-art hybrid accelerometers concept (b). Global maps plotted til d/o 90 (plots on the left); SH error spectra, log scale of the SH coefficient differences plotted til d/o 180 (plots on the right)}
\label{fig:GRV_map1}

\end{figure*}

\begin{figure*}
\begin{subcaptiongroup}
\begin{subfigure}{.5\textwidth}
    \centering
    \includegraphics[width=1\linewidth]{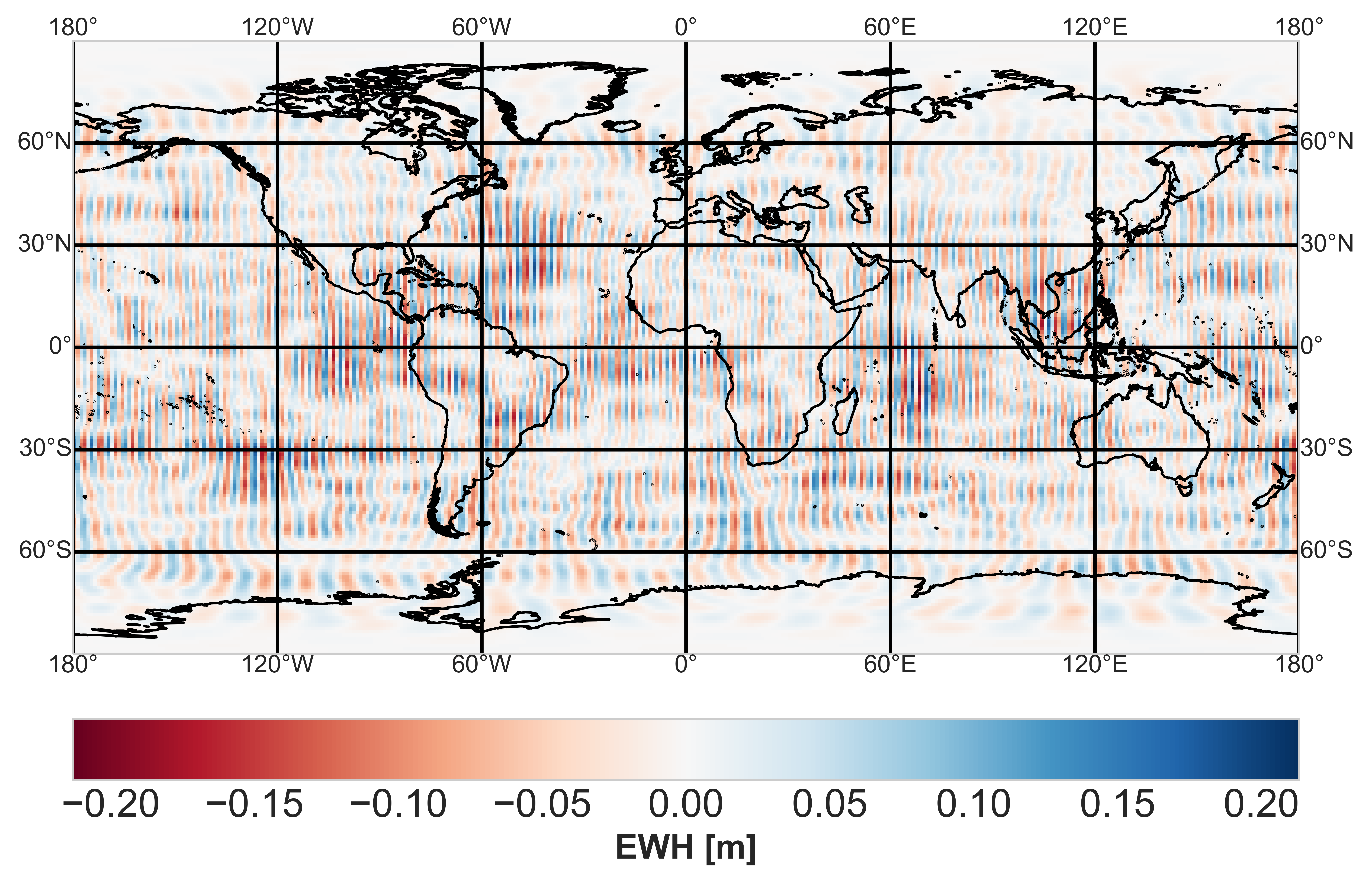}  
\end{subfigure}
\begin{subfigure}{.5\textwidth}
    \centering
    \includegraphics[width=1\linewidth]{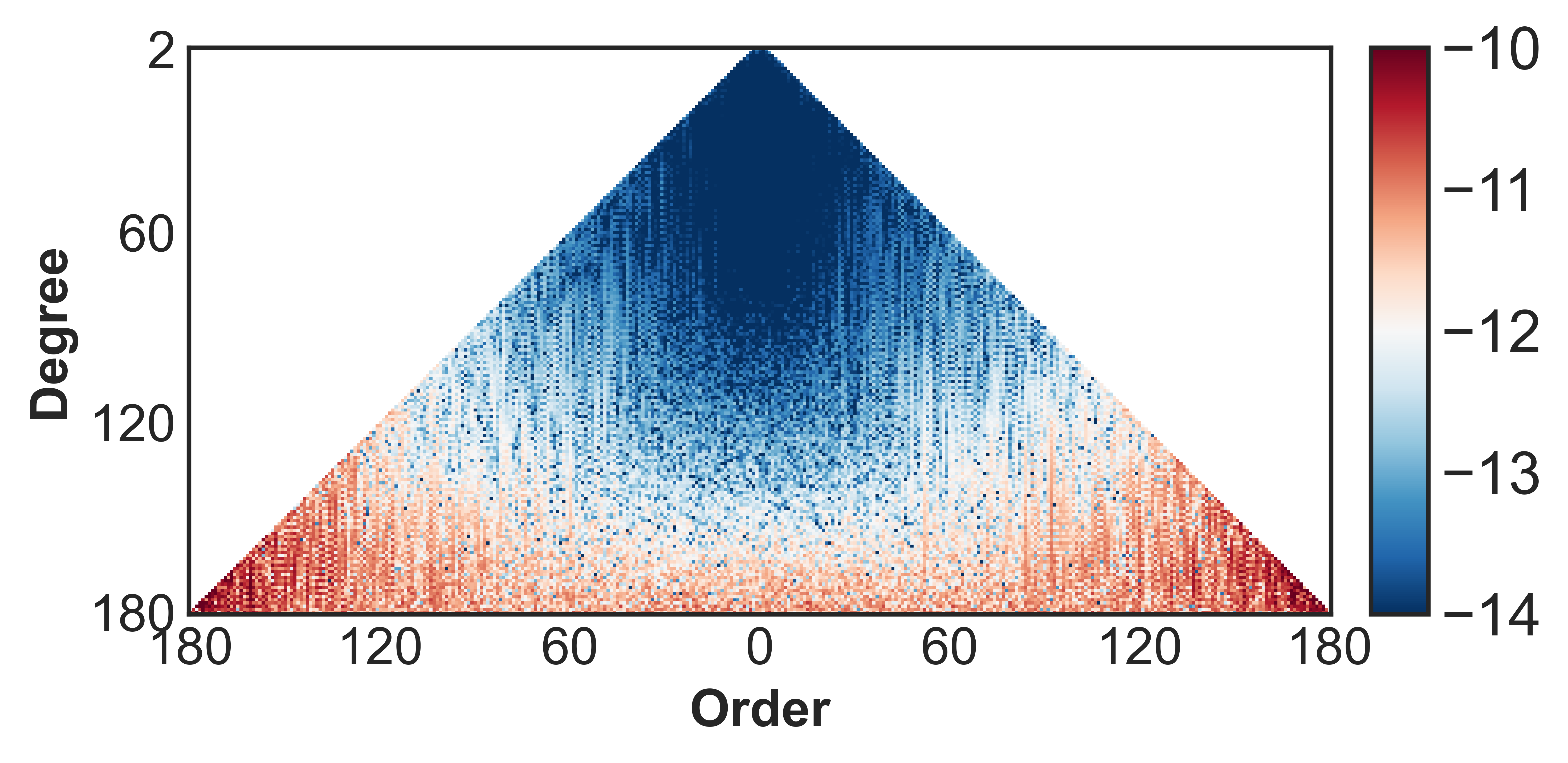}  
\end{subfigure}
\subcaption{Near-future Hybrid ACC $+$ LRI 2024}
\label{subfig:GRV_nearFuture_LRI24}
\begin{subfigure}{.5\textwidth}
    \centering
    \includegraphics[width=1\linewidth]{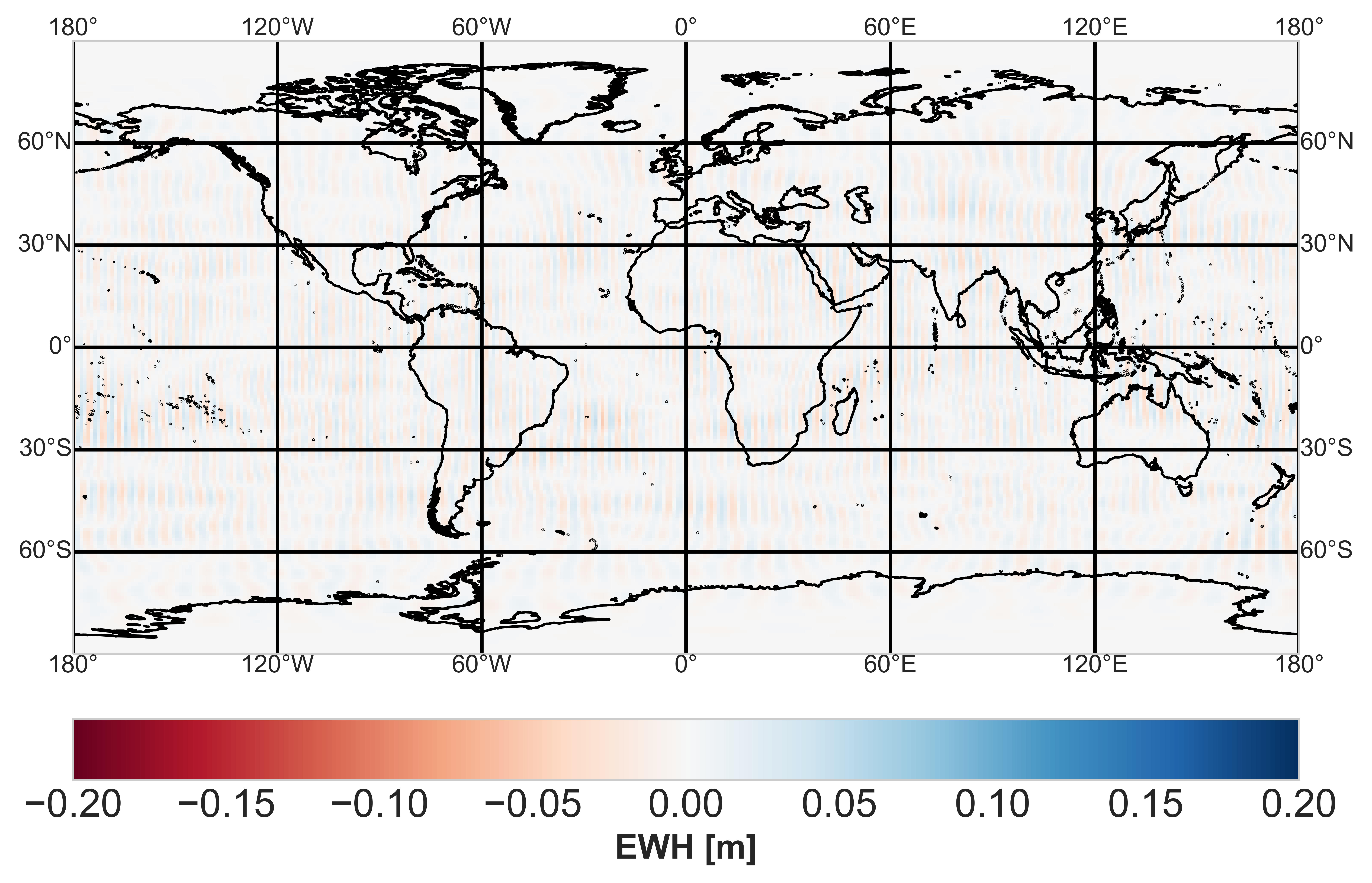}  
\end{subfigure}
\begin{subfigure}{.5\textwidth}
    \centering
    \includegraphics[width=1\linewidth]{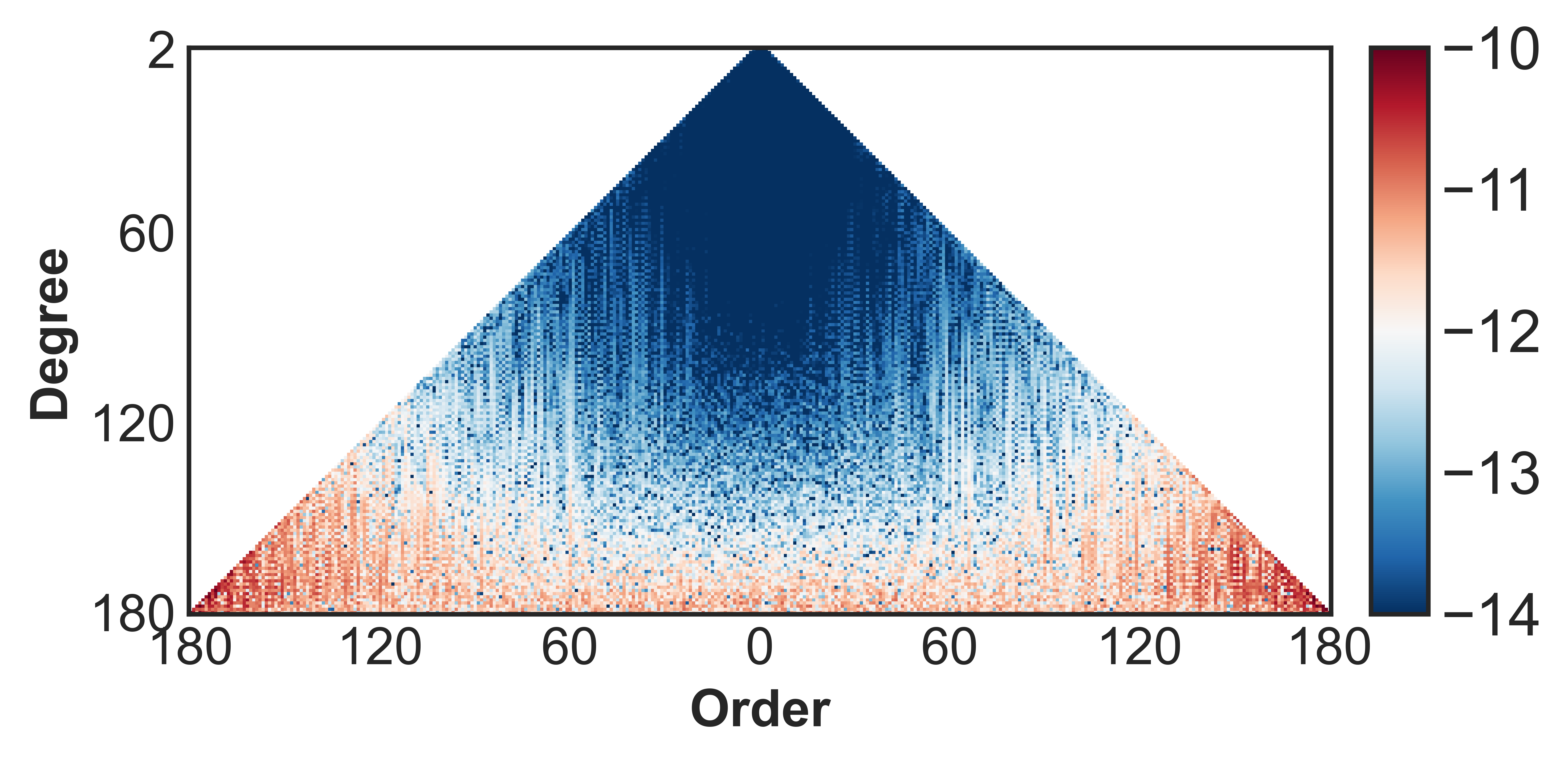}  
\end{subfigure}
\subcaption{Near-future Hybrid ACC $+$ LRI 2030}
\label{subfig:GRV_nearFuture_LRI30}
\end{subcaptiongroup}
\caption{Comparison of the recovered gravity fields retrieved til d/o 180 (without post-processing and filtering) from the near-future hybrid accelerometers concepts w.r.t. EIGEN 6C4. Global maps plotted til d/o 90 (plots on the left); SH error spectra, log scale of the SH coefficient differences plotted til d/o 180 (plots on the right)}
\label{fig:GRV_map2}

\end{figure*}

Figure~\ref{fig:gravRec} shows the degree RMS of the 
coefficient differences between reference and recovered gravity field in units of geoid height.
One can see one and two orders of improvement in geoid height by replacing the classical accelerometer with the state-of-the-art hybrid and near-future hybrid accelerometers, respectively. Figures~\ref{fig:GRV_map1} and~\ref{fig:GRV_map2} compare the global map of the residuals and the spherical harmonic error spectrum of the recovered gravity fields retrieved until degree and order (d/o) $180$ for the state-of-the-art and near-future hybrid accelerometers concepts with respect to EIGEN-6C4. 
Global maps are shown before post-processing and filtering, illustrating the improvements achieved by adding the Q-ACC and introducing the hybrid concept in the state-of-the-art scenario (see figure~\ref{subfig:GRV_stateArt}). Moreover, the produced global gravity field maps show a considerable reduction of the instrumental contribution to the striping effect after introducing the hybrid accelerometers. (see figure~\ref{subfig:GRV_nearFuture_LRI24}). The possibilities and challenges of having an improved Q-ACC for the near future are discussed in~\cite{hosseiniarani2024}. Finally, figure \ref{subfig:GRV_nearFuture_LRI30} displays the improvements in the gravity field that can be achieved by adding the LRI 2030 to the near future-hybrid accelerometer concept.

\section{Conclusions}
We considered an advanced Kalman-filter-based hybridization of electrostatic and quantum accelerometers by implementing a comprehensive noise model for satellite-based quantum sensors and considering the full impact of satellite rotation on the sensor's measurements. We studied the hybridization of state-of-the-art and near-future classical and quantum sensors and quantified their impacts on the performance of future satellite gravity missions. 

The filter converges in all scenarios. The solution benefis from the performances of the E-ACC measurements at higher frequencies, together with the high stability of CAI measurements at lower frequencies Thus the capability to online estimate and correct biases of the E-ACC. This holds true, even in presence of rotational effects. When a proper rotation compensation method is applied, the additional noise due to the rotation rate does not prevent the convergence of the filter. Our results indicate that implementing a hybrid accelerometer onboard a future gravity mission will improve the gravity field solution by one to two orders in both, lower and higher degrees. 
The global gravity field maps show considerable improvement in reducing the instrumental contribution to the striping effect by introducing hybrid accelerometers. Note that in this study, we only focused on the performance of the used sensors and we did not consider limitations due to insufficient knowledge of the background models.

\section*{Acknowledgments}
This work is supported by the Deutsche Forschungsgemeinschaft (DFG, German Research Foundation) Collaborative Research Center 1464 “TerraQ” – 434617780 and Germany’s Excellence Strategy – EXC-2123 “QuantumFrontiers” – 390837967, and by the European Union’s Horizon Europe research and innovation programme under grant agreement No 101081775 (CARIOQA-PMP project). This study is also partially supported by SpaceQNav project funded by the Federal Ministry for Economic Affairs and Climate Action (BMWK), Project 50NA2310A.
QB and FP acknowledge the support of a government grant managed by the Agence Nationale de la Recherche under the Plan France 2030, with the reference “ANR-22-PETQ-0005.”
B.T. acknowledges support from the Federal Ministry for Economic Affairs and Climate Action (BMWK), Project 50NA2106. J.M. and A.K. acknowledge support by Deutsches Zentrum für Luft- und Raumfahrt e.V. (DLR) for the project Q-BAGS.

\bibliography{refs}

@Article{Abich2019,
 author  = {Abich, Klaus and Abramovici, Alexander and Amparan, Bengie and Baatzsch, Andreas and Okihiro, Brian Bachman and Barr, David C. and others},
  title   = {In-orbit performance of the {GRACE} {Follow-On} {Laser} {Ranging} {Interferometer}},
  eid     = {031101},
  doi    = {10.1103/PhysRevLett.123.031101},
  number  = {3},
  volume  = {123},
  journal = {Phys Rev Lett},
  year    = {2019},
}

@article{Abrykosov2019,
title = {Impact of a novel hybrid accelerometer on satellite gravimetry performance},
journal = {Adv Space Res},
volume = {63},
number = {10},
pages = {3235-3248},
year = {2019},
issn = {0273-1177},
doi = {10.1016/j.asr.2019.01.034},
author = {Petro Abrykosov and Roland Pail and Thomas Gruber and Nassim Zahzam and Alexandre Bresson and Emilie Hardy and Bruno Christophe and Yannick Bidel and Olivier Carraz and Christian Siemes},
}

@article{Asenbaum2020,
  title = {Atom-Interferometric Test of the Equivalence Principle at the ${10}^{\ensuremath{-}12}$ Level},
  author = {Asenbaum, Peter and Overstreet, Chris and Kim, Minjeong and Curti, Joseph and Kasevich, Mark A.},
  journal = {Phys. Rev. Lett.},
  volume = {125},
  issue = {19},
  pages = {191101},
  numpages = {5},
  year = {2020},
  month = {Nov},
  publisher = {American Physical Society},
  doi = {10.1103/PhysRevLett.125.191101},
}

@article{Beaufils2023,
 author = {Beaufils, Quentin and Lefebve, Julien and Baptista, Joel Gomes and Piccon, Rapha{\"e}l and Cambier, Valentin and Sidorenkov, Leonid A. and Fallet, Christine and L{\'e}v{\`e}que, Thomas and Merlet, S{\'e}bastien and {Pereira dos Santos}, Franck},
 year = {2023},
 title = {Rotation related systematic effects in a cold atom interferometer onboard a Nadir pointing satellite},
 eid = {53},
 volume = {9},
 number = {1},
 journal = {NPJ microgravity},
 doi = {10.1038/s41526-023-00297-w},
}

@InProceedings{Christophe2018,
  author    = {Christophe, B. and Foulon, B. and Liorzou, F. and Lebat, V. and Boulanger, D. and Huynh, P.-A. and Zahzam, N. and Bidel, Y. and Bresson, A.},
  booktitle = {International {Symposium} on {Advancing} {Geodesy} in a {Changing} {World}},
  title     = {Status of development of the future accelerometers for {Next} {Generation} {Gravity} {Missions}},
  editor    = {Freymueller, Jeffrey T. and Sánchez, Laura},
  doi      = {10.1007/1345\_2018\_42},
  pages     = {85--89},
  publisher = {Springer International Publishing},
  series    = {International {Association} of {Geodesy} {Symposia}},
  volume    = {149},
  address   = {Cham},
  year      = {2018},
}

@Article{Flury2008,
  author  = {Jakob Flury and Srinivas Bettadpur and Byron D. Tapley},
  title   = {Precise accelerometry onboard the {GRACE} gravity field satellite mission},
  doi    = {10.1016/j.asr.2008.05.004},
  issue  = {8},
  pages   = {1414--1423},
  volume  = {42},
  journal = {Adv Space Res},
  year    = {2008},
}

@Article{Freier2016,
  author  = {Freier, Christian and Hauth, M and Schkolnik, V and Leykauf, B and Schilling, Manuel and Wziontek, H and Scherneck, H-G and Müller, J and Peters, A},
  title   = {Mobile quantum gravity sensor with unprecedented stability},
  eid     = {012050},
  doi    = {10.1088/1742-6596/723/1/012050},
  volume  = {723},
  journal = {J Phys: Conf Ser},
  year    = {2016},
}

@article{Geiger2020,
    author = {Geiger, Remi and Landragin, Arnaud and Merlet, Sébastien and Pereira Dos Santos, Franck},
    title = "{High-accuracy inertial measurements with cold-atom sensors}",
    journal = {AVS Quantum Science},
    volume = {2},
    issue = {2},
    pages = {024702},
    year = {2020},
    doi = {10.1116/5.0009093},
}

@Article{Haagmans2020,
  author  = {Haagmans, Roger and Siemes, Christian and Massotti, Luca and Carraz, Olivier and Silvestrin, Pierluigi},
  title   = {{ESA}’s next-generation gravity mission concepts},
  doi    = {10.1007/s12210-020-00875-0},
  pages   = {15--25},
  volume  = {31},
  journal = {Rend Lincei-Sci Fis},
  year    = {2020},
}

@inproceedings{hosseiniarani2022,
	title = {Kalman-filter Based Hybridization of Classic and Cold Atom Interferometry Accelerometers for Future Satellite Gravity Missions},
	author = {HosseiniArani, A. and Tennstedt, B. and Schilling, M. and Knabe, A. and Wu, Hu and Sch\"on, S. and M\"uller, J.},
	 doi = {10.1007/1345\_2022\_172},
  crossref = {Freymueller2022},
  pages = {221--231}
}

@article{Humphrey2023,
 author = {Humphrey, Vincent and Rodell, Matthew and Eicker, Annette},
 year = {2023},
 title = {Using Satellite-Based Terrestrial Water Storage Data: A Review},
 pages = {1489--1517},
 volume = {44},
 number = {5},
 journal = {Surveys in Geophysics},
 doi = {10.1007/s10712-022-09754-9},
}

@article{Kasevich1991,
  title = {Atomic interferometry using stimulated Raman transitions},
  author = {Kasevich, Mark and Chu, Steven},
  journal = {Phys. Rev. Lett.},
  volume = {67},
  issue = {2},
  pages = {181--184},
  year = {1991},
  publisher = {American Physical Society},
  doi = {10.1103/PhysRevLett.67.181},
}

@article{Lan2012,
 author = {Lan, Shau-Yu and Kuan, Pei-Chen and Estey, Brian and Haslinger, Philipp and M{\"u}ller, Holger},
 year = {2012},
 title = {Influence of the Coriolis force in atom interferometry},
 eid = {090402},
 volume = {108},
 issue = {9},
 journal = {Phys Rev Lett},
 doi = {10.1103/PhysRevLett.108.090402},
}

@article{Lecomte2023,
 author = {Lecomte, Hugo and Rosat, S{\'e}verine and Mandea, Mioara and Dumberry, Mathieu},
 year = {2023},
 title = {Gravitational Constraints on the {Earth}'s Inner Core Differential Rotation},
 pages = {e2023GL104790},
 volume = {50},
 issue = {23},
 journal = {Geophysical Research Letters},
 doi = {10.1029/2023GL104790}
}

@Article{Leveque2021,
  author  = {L{\'e}v{\`e}que, T. and Fallet, C. and Mandea, M. and Biancale, R. and Lemoine, J. M and Tardivel, S. and Delavault, S. and Piquereau, A. and Bourgogne, S. and Pereira Dos Santos, F. and Battelier, B. and Bouyer, Ph.},
  title   = {Gravity Field Mapping Using Laser Coupled Quantum Accelerometers in Space},
  doi    = {10.1007/s00190-020-01462-9},
  eid  = {15},
  volume  = {95},
  journal = {J Geod},
  year    = {2021}
}

@article{leveque2009,
  title={Enhancing the area of a Raman atom interferometer using a versatile double-diffraction technique},
  author={L{\'e}v{\`e}que, Thomas and Gauguet, Alexandre and Michaud, Franck and Dos Santos, F Pereira and Landragin, Arnaud},
  journal={Physical review letters},
  volume={103},
  number={8},
  pages={080405},
  year={2009},
  publisher={APS}
}

@article{Migliaccio2019,
 author = {Migliaccio, Federica and Reguzzoni, Mirko and Batsukh, Khulan and Tino, Guglielmo Maria and Rosi, Gabriele and Sorrentino, Fiodor and Braitenberg, Carla and Pivetta, Tommaso and Barbolla, Dora Francesca and Zoffoli, Simona},
 year = {2019},
 title = {{MOCASS}: A Satellite Mission Concept Using Cold Atom Interferometry for Measuring the {Earth} Gravity Field},
 pages = {1029--1053},
 volume = {40},
 number = {5},
 journal = {Surv Geophys},
 doi = {10.1007/s10712-019-09566-4},
}

@Article{Pail2015,
  author  = {Pail, Roland and Bingham, Rory and Braitenberg, Carla and Dobslaw, Henryk and Eicker, Annette and Güntner, Andreas and Horwath, Martin and Ivins, Eric and Longuevergne, Laurent and Panet, Isabelle and Wouters, Bert and {IUGG Expert Panel}},
  title   = {Science and user needs for observing global mass transport to understand global change and to benefit society},
  doi    = {10.1007/s10712-015-9348-9},
  number  = {6},
  pages   = {743--772},
  volume  = {36},
  journal = {Surv Geophys},
  year    = {2015},
}

@article{Scanlon2023,
 author = {Scanlon, Bridget R. and Fakhreddine, Sarah and Rateb, Ashraf and de Graaf, Inge and Famiglietti, Jay and Gleeson, Tom and Grafton, R. Quentin and Jobbagy, Esteban and Kebede, Seifu and Kolusu, Seshagiri Rao and Konikow, Leonard F. and {Di Long} and Mekonnen, Mesfin and Schmied, Hannes M{\"u}ller and Mukherjee, Abhijit and MacDonald, Alan and Reedy, Robert C. and Shamsudduha, Mohammad and Simmons, Craig T. and Sun, Alex and Taylor, Richard G. and Villholth, Karen G. and V{\"o}r{\"o}smarty, Charles J. and Zheng, Chunmiao},
 year = {2023},
 title = {Global water resources and the role of groundwater in a resilient water future},
 pages = {87--101},
 volume = {4},
 number = {2},
 journal = {Nature Reviews Earth {\&} Environment},
 doi = {10.1038/s43017-022-00378-6},
}

@Article{Tapley2019,
  author  = {Tapley, Byron D. and Watkins, Michael M. and Flechtner, Frank and Reigber, Christoph and Bettadpur, Srinivas and Rodell, Matthew and others},
  title   = {Contributions of {GRACE} to understanding climate change},
  doi    = {10.1038/s41558-019-0456-2},
  number  = {5},
  pages   = {358--369},
  volume  = {9},
  journal = {Nat Clim Change},
  year    = {2019},
}

@Article{Christophe2015,
  author  = {Christophe, B. and Boulanger, D. and Foulon, B. and Huynh, P. -A. and Lebat, V. and Liorzou, F. and Perrot, E.},
  title   = {A new generation of ultra-sensitive electrostatic accelerometers for {GRACE} {Follow}-on and towards the next generation gravity missions},
  doi    = {10.1016/j.actaastro.2015.06.021},
  pages   = {1--7},
  volume  = {117},
  journal = {Acta Astronaut},
  year    = {2015},
}

@Article{Schilling2020,
  author  = {Schilling, Manuel and Wodey, \'{E} and Timmen, Ludger and Tell, Dorothee and Zipfel, Klaus H. and Schlippert, Dennis and Schubert, Christian and Rasel, Ernst M. and Müller, Jürgen},
  title   = {Gravity field modelling for the {Hannover} 10 m atom interferometer},
  eid     = {122},
  doi    = {10.1007/s00190-020-01451-y},
  number  = {12},
  volume  = {94},
  journal = {J Geod},
  year    = {2020},
}

@article{Shihora2022,
 author = {Shihora, Linus and Balidakis, Kyriakos and Dill, Robert and Dahle, Christoph and Ghobadi--Far, Khosro and Bonin, Jennifer and Dobslaw, Henryk},
 year = {2022},
 title = {Non--Tidal Background Modeling for Satellite Gravimetry Based on Operational ECWMF and ERA5 Reanalysis Data: AOD1B RL07},
 volume = {127},
 issue = {8},
 issn = {2169-9313},
 journal = {Journal of Geophysical Research: Solid Earth},
 doi = {10.1029/2022JB024360},
}

@Article{Loomis2020,
  author  = {Loomis, Bryant D. and Rachlin, Kenneth E. and Wiese, David N. and Landerer, Felix W. and Luthcke, Scott B.},
  title   = {Replacing {GRACE}/{GRACE}-{FO} with {Satellite} {Laser} {Ranging}: impacts on {Antarctic} ice sheet mass change},
  doi    = {10.1029/2019GL085488},
  number  = {3},
  pages   = {e2019GL085488},
  volume  = {47},
  journal = {Geophys Res Lett},
  year    = {2020},
}

@article{Chen2022,
 author = {Chen, Jianli and Cazenave, Anny and Dahle, Christoph and Llovel, William and Panet, Isabelle and Pfeffer, Julia and Moreira, Lorena},
 year = {2022},
 title = {Applications and Challenges of GRACE and GRACE Follow-On Satellite Gravimetry},
 pages = {305--345},
 volume = {43},
 number = {1},
 journal = {Surveys in Geophysics},
 doi = {10.1007/s10712-021-09685-x},
}

@Article{Mandea2020,
  author  = {Mioara Mandea and V{\'{e}}ronique Dehant and Anny Cazenave},
  title   = {{GRACE}--Gravity Data for Understanding the Deep {Earth's} Interior},
  eid     = {4186},
  doi    = {10.3390/rs12244186},
  issue  = {24},
  volume  = {12},
  journal = {Remote Sens},
  year    = {2020},
}

@inproceedings{Marque2010,
  title={Accelerometers of the {GOCE} mission: return of experience from one year of in-orbit},
  author={Marque, Jean-Pierre and Christophe, Bruno and Foulon, Bernard},
  booktitle={ESA Living Planet Symposium, 28.06.-02.07.2010, Bergen, Norway},
  volume={686},
  pages={57},
  year={2010}
}

@inproceedings{Weddig2021,
	title = {Performance evaluation of a three-dimensional cold atom interferometer based inertial navigation system},
	isbn = {978-1-66543-178-1},
	 doi = {10.1109/ISS52949.2021.9619776},
	booktitle = {2021 {DGON} {Inertial} {Sensors} and {Systems} ({ISS}), 29.-30.09.2021 Braunschweig, Germany},
	publisher = {IEEE},
	author = {Weddig, N. B. and Tennstedt, B. and Sch\"on, S.},
	year = {2021},
	pages = {1--20},
	editor = {Hecker, P.}
}

@Article{Woeske2019,
  author  = {Florian Wöske and Takahiro Kato and Benny Rievers and Meike List},
  title   = {{GRACE} accelerometer calibration by high precision non-gravitational force modeling},
  doi    = {10.1016/j.asr.2018.10.025},
  issue  = {3},
  pages   = {1318-1335},
  volume  = {63},
  journal = {Adv Space Res},
  year    = {2019},
}

@article{Wiese2022,
 author = {Wiese, D. N. and Bienstock, B. and Blackwood, C. and Chrone, J. and Loomis, B. D. and Sauber, J. and Rodell, M. and Baize, R. and Bearden, D. and Case, K. and Horner, S. and Luthcke, S. and Reager, J. T. and Srinivasan, M. and Tsaoussi, L. and Webb, F. and Whitehurst, A. and Zlotnicki, V.},
 year = {2022},
 title = {The Mass Change Designated Observable Study: Overview and Results},
 pages = {e2022EA002311},
 volume = {9},
 issue = {8},
 issn = {2333-5084},
 journal = {Earth and Space Science},
 doi = {10.1029/2022EA002311},
}

@phdthesis{Wu2016,
	address = {Hannover},
	type = {Dissertation},
	title = {Gravity field recovery from {GOCE} observations},
	school = {Leibniz Universität Hannover},
	author = {Wu, Hu},
	year = {2016},
  }

@article{Zahzam2022,
 author = {Zahzam, Nassim and Christophe, Bruno and Lebat, Vincent and Hardy, Emilie and Huynh, Phuong-Anh and Marquet, No{\'e}mie and Blanchard, C{\'e}dric and Bidel, Yannick and Bresson, Alexandre and Abrykosov, Petro and Gruber, Thomas and Pail, Roland and Daras, Ilias and Carraz, Olivier},
 year = {2022},
 title = {Hybrid electrostatic-atomic accelerometer for future space gravity missions},
 volume = {14},
 issue = {14},
 journal = {Remote Sensing},
 doi = {10.3390/rs14143273},
}

@inproceedings{Knabe2022,
 author = {Knabe, Annike and Schilling, Manuel and Wu, Hu and HosseiniArani, Alireza and M{\"u}ller, J{\"u}rgen and Beaufils, Quentin and {Pereira dos Santos}, Franck},
 title = {{The Benefit of Accelerometers Based on Cold Atom Interferometry for Future Satellite Gravity Missions}},
 doi = {10.1007/1345\_2022\_151},
  crossref = {Freymueller2022},
 pages = {213--220},
}

@Proceedings{Freymueller2022,
  editor = {Freymueller, J.T. and Sanchez, L.},
  series = {International Association of Geodesy Symposia},
  volume    = {154},
  booktitle = {Geodesy for a Sustainable Earth},
  year      = {2022},
  address = {Cham},
  publisher = {Springer},
}

@article{hosseiniarani2024,
      title={Advances in Atom Interferometry and their Impacts on the Performance of Quantum Accelerometers On-board Future Satellite Gravity Missions}, 
      author={Alireza HosseiniArani and Manuel Schilling and Quentin Beaufils and Annike Knabe and Benjamin Tennstedt and Alexey Kupriyanov and Steffen Schön and Franck Pereira dos Santos and Jürgen Müller},
      year={2024},
      eprint={2404.10471},
      archivePrefix={arXiv},
      primaryClass={physics.ins-det}
}

@article{KUPRIYANOV2024,
title = {Benefit of enhanced electrostatic and optical accelerometry for future gravimetry missions},
journal = {Advances in Space Research},
volume = {73},
issue = {6},
pages = {3345-3362},
year = {2024},
issn = {0273-1177},
doi = {10.1016/j.asr.2023.12.067},
author = {Alexey Kupriyanov and Arthur Reis and Manuel Schilling and Vitali Müller and Jürgen Müller},
}

@article{tennstedt2023atom,
  title={Atom Strapdown: Toward Integrated Quantum Inertial Navigation Systems},
  author={Tennstedt, Benjamin and Rajagopalan, Ashwin and Weddig, Nicolai B and Abend, Sven and Sch{\"o}n, Steffen and Rasel, Ernst M},
  journal={NAVIGATION: Journal of the Institute of Navigation},
  volume={70},
issn   = {0028-1522},
pages  = {navi.604},
  issue={4},
  year={2023},
  publisher={Institute of Navigation}
}

@inproceedings{Woske.2016,
  title={{Development of a high precision simulation tool for gravity recovery missions like GRACE}},
  author={W{\"o}ske, Florian and Kato, T and List, M and Rievers, B},
  booktitle={Proceedings of the 26th AAS/AIAA space flight mechanics meeting held February},
  editor = {Zanetti, R and Russel, R.P. and Ozimek, M.T. and Bowes, A.L.},
  volume={158},
  pages={2445-2457},
  year={2016}
}

@article{Woske.2018,
 author = {W{\"o}ske, Florian and Kato, Takahiro and Rievers, Benny and List, Meike},
 title = {{GRACE accelerometer calibration by high precision non-gravitational force modeling}},
 urldate = {2020-05-03},
 pages = {1318--1335},
 volume = {63},
 number = {3},
 issn = {02731177},
 journal = {Advances in Space Research},
 doi = {10.1016/j.asr.2018.10.025},
 year={2018}
}

\clearpage

\end{document}